\documentclass[a4paper,12pt]{article}
\usepackage{a4wide}
\usepackage{amsmath}
\usepackage{graphicx}

\newcommand\be{\begin{equation}}
\newcommand\ee[1]{\label{#1}\end{equation}}
\newcommand\ie{\textit{i.e.}}

\renewcommand\Re{\mathrm{Re}}
\newcommand\etal{\textit{et al.}}

\begin{document}
\title{Universality of the turbulent velocity profile}
\author{Paolo Luchini\\
\small Dipartimento di Ingegneria Industriale \\\small  Universit\`a di Salerno\\\small  84084 Fisciano, Italy\\
\small luchini{@}unisa.it}

\maketitle

For nearly a century the universal logarithmic behaviour of the mean velocity profile in a parallel flow was a mainstay of turbulent fluid mechanics and its teaching \cite{Panton,White}. Yet many experiments and numerical simulations are not fit exceedingly well by it, and the question whether the logarithmic law is indeed universal keeps turning up in discussion and in writing \cite{Barenblatt,Chauhan}. Large experiments have been set up in different parts of the world to confirm or deny the logarithmic law and accurately estimate von K\'arm\'an's constant, the coefficient that governs it \cite{superpipe,ciclope}. We show that the discrepancy among flows in different (circular or plane) geometries, and between these and the logarithmic law, can be ascribed to the effect of the pressure gradient. When this effect is accounted for in the form of a higher-order perturbation, universal agreement and a satisfactorily simple formulation are recovered.
\newpage

Turbulent flow in a parallel channel or duct is the prototype problem of wall-bounded turbulence. Be it between parallel plane walls or in a circular pipe, this is the contained turbulent flow with highest geometrical symmetry, and also the one whose physical properties are most studied. In the ideal case of infinite extension in the other directions, the mean velocity is directed parallel to the walls and is a function of the wall-normal coordinate only, namely it is a function of a single variable known as the velocity profile $u(z)$.

Although the equations governing the time-evolution of the flow are known, the Navier--Stokes equations, the only workable way to extract its time average is to actually run  a time-resolved simulation for a long time and take its average (so called Direct Numerical Simulation, or DNS), equivalent to running an experiment for a long enough time and taking its average.
In the case of a parallel flow there are so few parameters that a universal shape of the velocity profile can be identified. The classical route to do so is based on dimensional analysis coupled with a few critical assumptions.

Almost a century ago, Prandtl \cite{Prandtl} recognized by his mixing-length argument that the velocity profile in a duct or pipe would have to be approximately logarithmic. The theory was then given its present form based on dimensional analysis by Millikan \cite{Millikan}.
The physical parameters affecting the phenomenon, in the case of an incompressible newtonian fluid like water or air, are the density $\rho$ and kinematic viscosity $\nu$ of the fluid, a typical dimension $h$ of the container (which we shall assume to be the distance from the wall to the symmetry axis, \ie\ the radius of a pipe or the semi-distance of two parallel plane walls), and the externally imposed pressure gradient $p_x$. The latter is bound to the mean shear stress $\tau_w$ exerted on the container walls by a simple force balance: if $A$ is the area of the duct's cross-section and $LP$ its lateral area, product of length $L$ and perimeter $P$, static equilibrium requires that $-A p_x L=LP\tau_w$, or
\be
p_x=-4\tau_w/D_H,
\ee{balance}
where the quantity $D_H=4A/P$ goes under the name of hydraulic diameter (because in the case of a circular pipe it coincides with its diameter, and in other cases it has the role of diameter of an equivalent pipe). Therefore, for a given geometry only one of $\tau_w$ or $p_x$ must be specified; nevertheless these two quantities have a very different physical meaning and in what follows the distinction will be crucial.

Using the dimensional parameters $\rho$, $\nu$, $h$, $\tau_w$ and the wall-normal coordinate $z$, at most two independent dimensionless groupings can be constructed. Typically a characteristic velocity is constructed first, as $u_\tau=\sqrt{\tau_w/\rho}$, and then a ``viscous length" $l=\nu/u_\tau$. The ratio $h/l$ is the Reynolds number $\Re_\tau=hu_\tau/\nu$, and it must be large or the flow would not be turbulent. The classical asymptotic theory of the turbulent velocity profile after Millikan is constructed in the limit of $\Re_\tau\rightarrow\infty$. The basic ansatz is that near the wall, in the region $z\ll h$ known as the ``wall layer", the velocity profile turns out independent of height $h$ and is thus a dimensionless function of a single variable,
\be
u^+=f(z^+),
\ee{wall}
where $u^+=u/u_\tau$ and $z^+=z/l=zu_\tau/\nu$. (Quantities denoted by a $^+$ are also commonly described as being measured in ``wall units".)

Conversely, for $z\gg l$ the velocity profile becomes independent of length $l$, or equivalently of the fluid's viscosity, but with a catch: because of galileian invariance, having lost the reference of a wall of whose motion we assume the flow has become independent, only velocity differences can be defined. Therefore the difference between the centerline velocity $U=u(h)$ and the generic velocity $u(z)$ is a function of the single variable $Z=z/h$ as
\be
U^+-u^+=F(Z).
\ee{defect}
$U-u$ is known as the velocity \emph{defect}, and the region $z\gg l$ as the ``defect layer".

Since by assumption $h\gg l$, there is a range of $z$ where the two conditions $z\gg l$ and $z\ll h$ can be simultaneously verified. We are then in the ``overlap layer", where the velocity profile is independent of both $l$ and $h$. Having exhausted the available dimensional quantities, only a dimensionless constant can be constructed in this layer. This is von K\'arm\'an's constant 
\be
\kappa=\frac{u_\tau}{zu_z}
\ee{vonK}
(involving the velocity derivative in place of the velocity because, just as in the defect layer, velocity can only be determined up to an additive constant). It follows by integration of \eqref{vonK} that in the overlap layer the velocity profile is logarithmic:
\be
u^+ = \kappa^{-1}\log(z^+)+B = \kappa^{-1}\log(Z)+C,
\ee{profile}
where $B$ and $C$ are suitable integration constants allowing a matching with the wall and defect layers, and tied to one another as $C=B+\log(h/l)=B+\log(\Re_\tau)$. Notice that out of this reasoning, while $C$ depends on geometry through $\Re_\tau$, the $\kappa$ and $B$ constants and the so called ``law of the wall" $f(z^+)$ must be universal.

\begin{figure}[h!]
\includegraphics{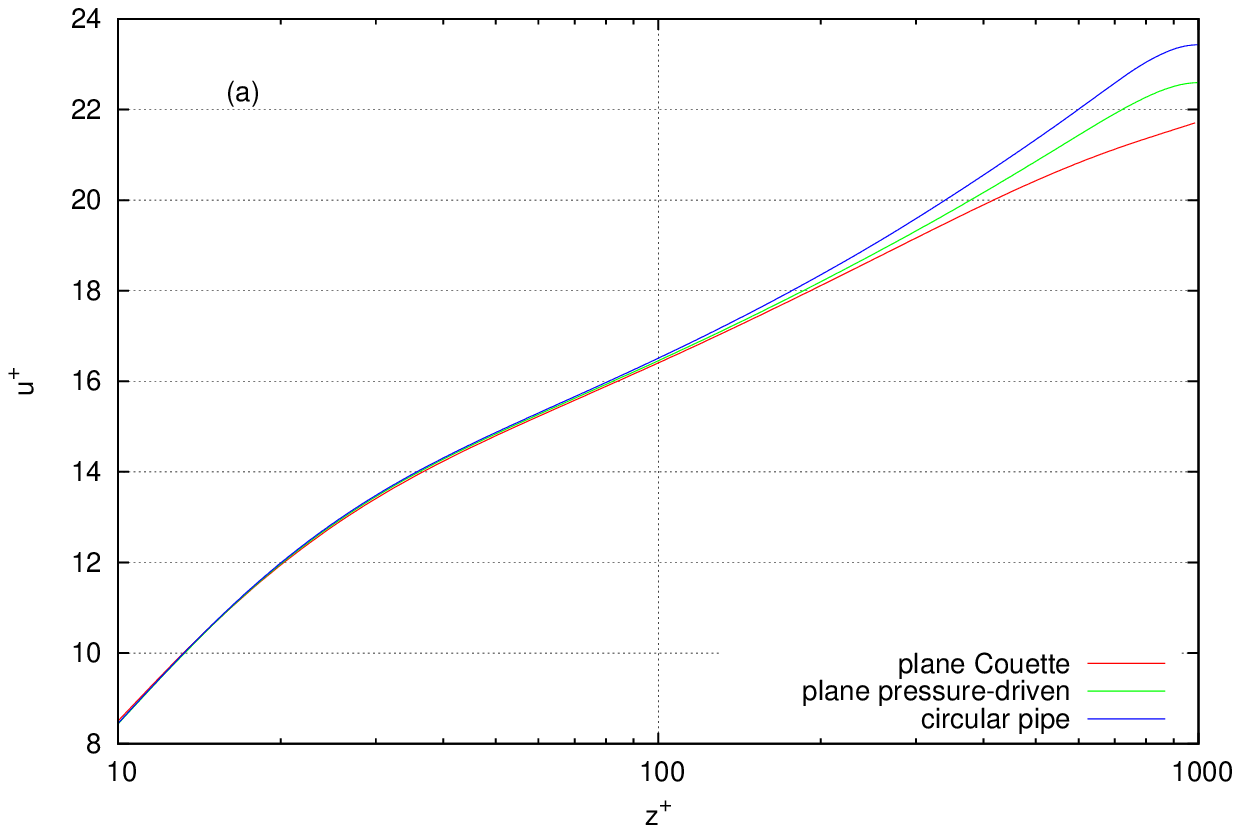}
 
\includegraphics{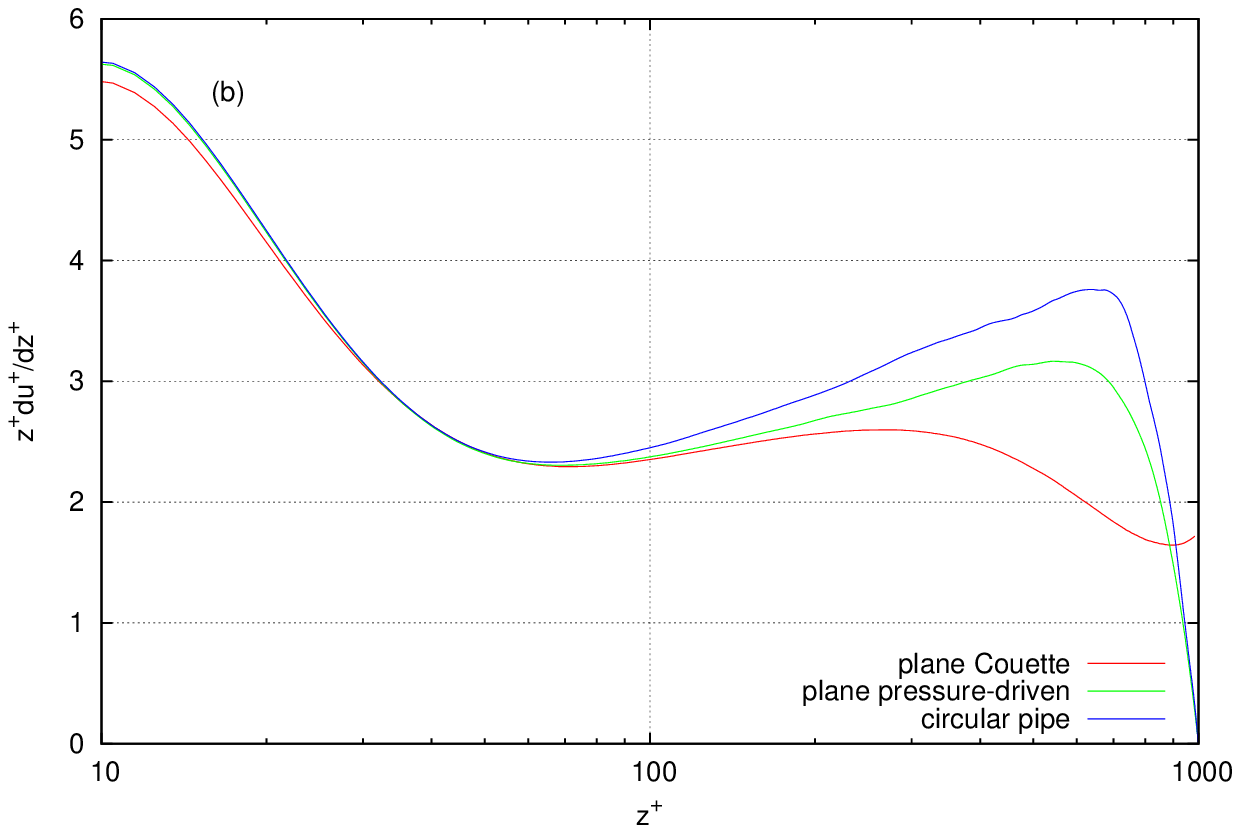}
\caption{Velocity profile in wall units versus wall-normal coordinate on a logarithmic scale (a) and its logarithmic derivative (b), for three different geometries at $\Re_\tau=1000$. DNS data for the circular pipe flow are taken from El Khoury \etal\ \cite{Schlatter}. DNS data for the pressure-driven plane parallel flow are taken from Lee \& Moser \cite{Moser}. DNS data for plane Couette flow are taken from Pirozzoli \etal\ \cite{Pirozzoli}.}
\label{threevel}
\end{figure}

The above theory has been the pillar of the description of wall-bounded turbulence for nearly a century, but there are problems. Whereas since the first experimental measurements of Nikuradse \cite{Nikuradse} it was clear that von K\'arm\'an's 
constant $\kappa$ is of the order of $0.4$, attempts to determine it more precisely from subsequent experiments and, once they became available, from DNS have proved elusive. A number of authors developed alternative explanations, most prominent being the incomplete-similarity theory of Barenblatt \cite{Barenblatt}, leading to a power law with a Reynolds-dependent exponent, and recently it was suggested \cite{Chauhan} that von K\'arm\'an's constant $\kappa$ is not universal but indeed it takes three different values for a channel with parallel plane walls, for a circular pipe and for the boundary layer (approximated as a parallel flow). The difficulty is exemplified in Fig. \ref{threevel}, which reports, at a particular Reynolds number $\Re_\tau=1000$, the numerical velocity profiles for circular-pipe, plane pressure-driven and plane Couette flow and their logarithmic derivatives, as obtained from DNS data available in the literature. Clearly a single set of logarithmic-law constants cannot fit all three. In addition, the velocity law doesn't appear especially close to being logarithmic, as may be seen even better from its derivative which ought to be constant if it was.

However, it should not be forgotten that the theory of the logarithmic layer is an asymptotic approximation for $\Re_\tau\rightarrow\infty$. In this sense every similarity theory is incomplete, but like all asymptotic approximations it can be improved with the addition of higher-order terms. We have seen that the influence of the physical parameters $\nu$ (viscosity), $h$ (geometry), and $p_x$ [pressure gradient, tied to $\tau_w$ and to geometry by the relationship \eqref{balance}] tends to be lost in the overlap layer; but, whereas viscosity is ``small" in the sense that $u_\tau z/\nu \gg 1$, and geometry is ``far" in the sense that $z\ll h$, the pressure gradient is uniform and acts all along the velocity profile. Therefore we can modify Millikan's ansatz of independence that led to \eqref{vonK} into the new statement:
\begin{quotation}
In the overlap layer the quantity $zu_z/u_\tau$ is \emph{almost} independent of $h$ and $\nu$, but is acted upon by $p_x$ as a \emph{small perturbation}.
\end{quotation}
In general the inclusion of an additional dimensional parameter ($p_x$) could mean that  $zu_z/u_\tau$ is no longer a constant $\kappa^{-1}$ but a new arbitrary function of a single variable. But, by specifying that $p_x$ is a small perturbation, we impose this function to be linear in $p_x$. Then there is only one dimensionally correct linear extension of \eqref{vonK}, and it is
\be
\frac{u_z}{u_\tau}=\frac{1}{\kappa z}-A_1\frac{p_x}{\tau_w},
\ee{newvonK}
whence, by integration, equation \eqref{profile} becomes
\be
u^+ = \kappa^{-1}\log(z^+)+A_1g\Re_\tau^{-1}z^+ +B = \kappa^{-1}\log(Z)+A_1 gZ +C.
\ee{newprofile}
Here $A_1$ is a new universal constant and the geometry parameter $g=-hp_x/\tau_w=4h/D_H$ takes the values of $g=2$ for circular pipe flow, $g=1$ for pressure-driven flow between steady plane walls and $g=0$ for turbulent Couette flow between countermoving plane walls. As may be seen the new term proportional to $p_x$ acts as a small perturbation of order $\Re_\tau^{-1}$ in the wall layer, consistent with the idea that (\ref{wall},\ref{profile}) were just the leading term of a universal asymptotic expansion in reciprocal powers of $\Re_\tau$, and becomes O$(1)$ in the defect layer where it merges smoothly with the function $F(Z)$ of \eqref{defect}, which itself is allowed to vary from one geometry to another. This is analogous to the method of matched asymptotic expansions developed by van Dyke \cite{vanDyke} for the laminar boundary layer: the first-order [O$(\Re_\tau^{-1})$] asymptotic correction to the inner solution becomes the first-order [O$(Z)$] term of the Taylor series for the outer solution.

\begin{figure}[h!]
\includegraphics{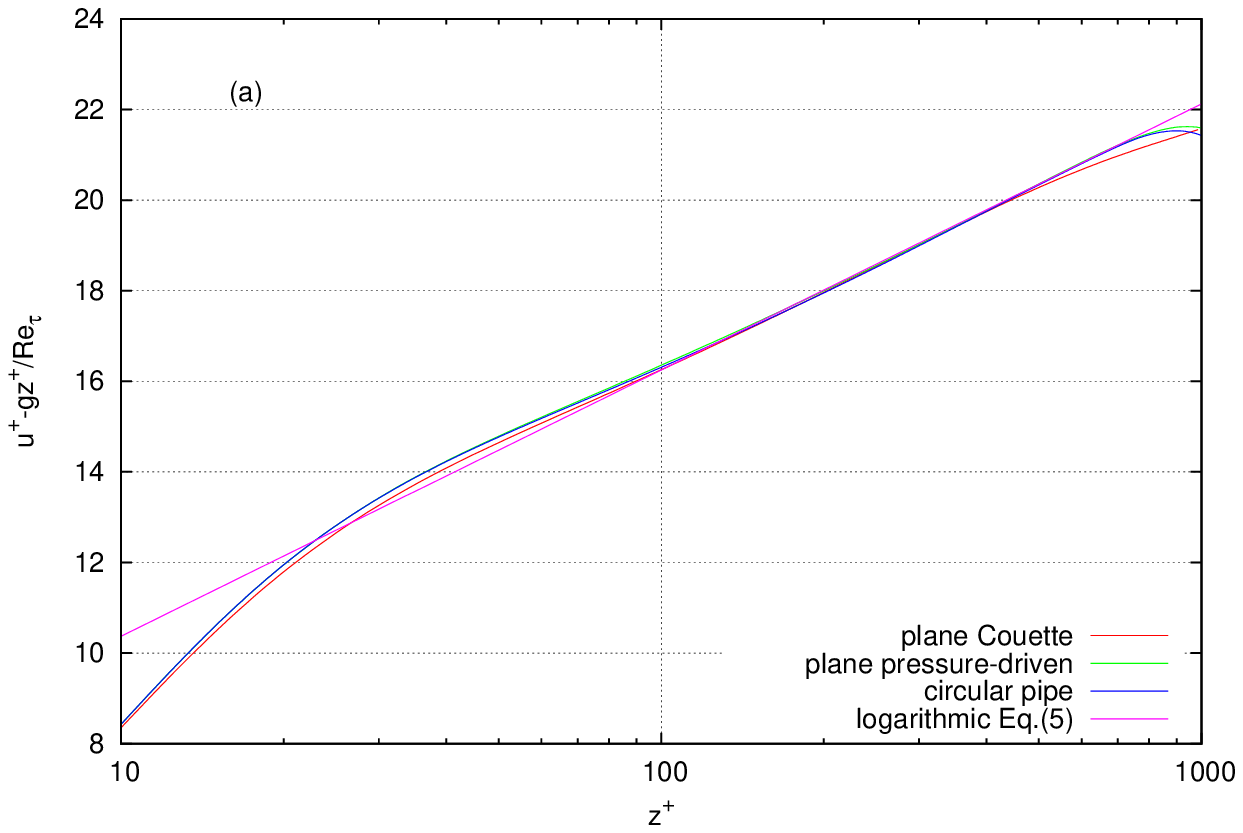}

\includegraphics{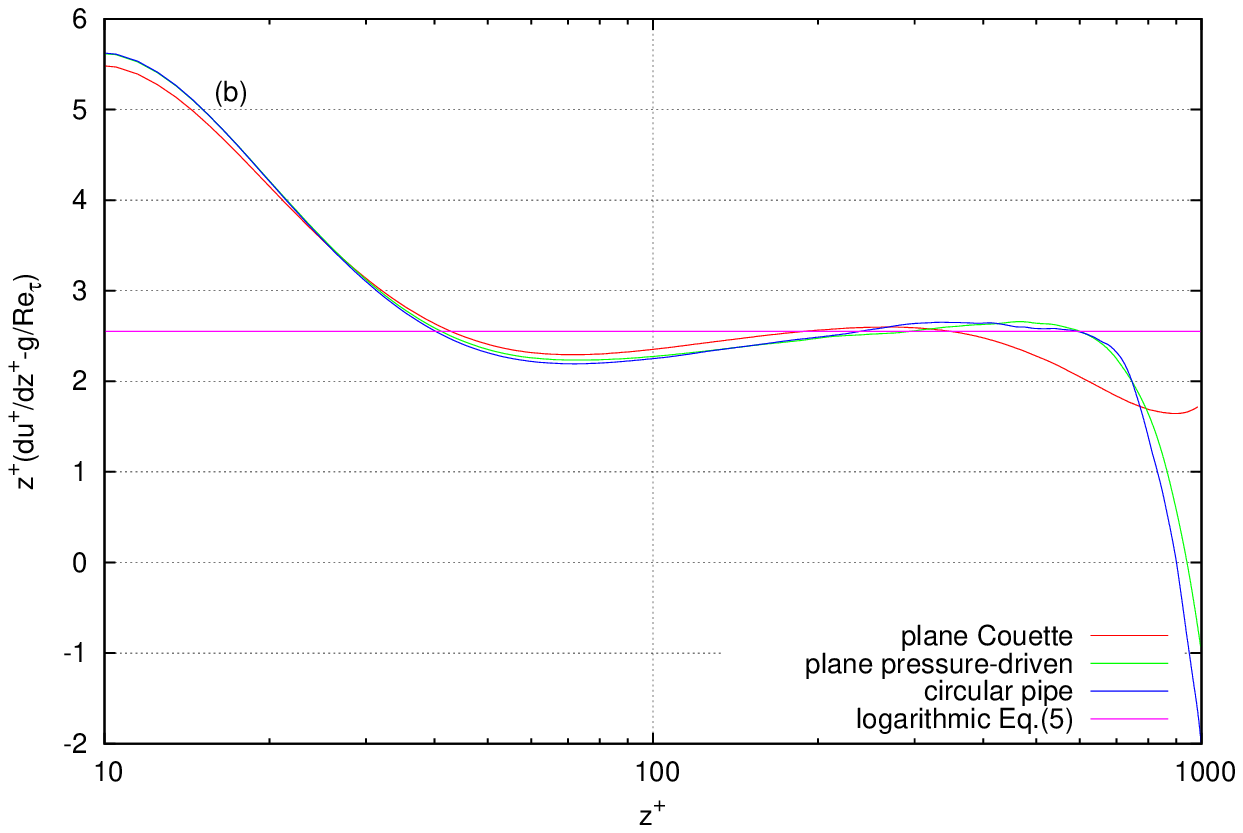}
\caption{Velocity profile (a) and its logarithmic derivative (b), after subtraction of the pressure-gradient term $gz^+/\Re_\tau$.}
\label{threevelpost}
\end{figure}

As the outcome of a fitting of available numerical and experimental data at several values of the Reynolds number, the details of which are given in \cite{inprep}, the value of $A_1$ comes close to unity, and it is tempting to set it exactly equal to 1. In fact a good overall fit of \eqref{newprofile} is provided by
\be
\kappa=0.392;\quad A_1=1; \quad B=4.50.
\ee{values}
In order to illustrate the effect of this modification, Fig. \ref{threevelpost} 
contains the same velocity profiles as Fig. \ref{threevel} each diminished of the first-order correction $A_1g\Re_\tau^{-1}z^+$; these profiles must coincide much more closely across the three cases if \eqref{newprofile} is correct. As can be seen this is indeed the case, and the common behaviour of the three curves is much closer to logarithmic than before in the central region. (The initial, non-logarithmic part of the curve up to about $z^+=250$ is the trace of the universal law of the wall \eqref{wall} with its own higher-order correction, but this is a longer story that requires a separate account \cite{inprep}.)

In conclusion, the logarithmic law of the turbulent velocity profile is indeed universal across different geometries of wall-bounded flow, provided the effect of pressure gradient is accounted for. The present equation \eqref{newprofile} does not contradict the classical equation \eqref{profile} but rather extends it with the inclusion of a higher-order term, and in fact tends to \eqref{profile} for $\Re_\tau\rightarrow\infty$. Nevertheless, this higher-order term is essential when using the logarithmic law for practical purposes or when estimating von K\'arm\'an's constant $\kappa$ from numerical or experimental data taken at practical values of the Reynolds number; omitting it can give the impression that the logarithmic law is invalid or that $\kappa$ is not universal.

\end{document}